\documentclass{article}
\usepackage{emulateapj}
\usepackage{graphicx}

\slugcomment{To Appear in ApJ}

\begin{document}

\title{Non-LTE Model Atmospheres for Late-Type Stars II. 
Restricted NLTE Calculations for a Solar-Like Atmosphere.}

\author{Carlos Allende Prieto}
\affil{McDonald Observatory and Department of Astronomy, University of Texas,
Austin, TX 78712}

\author{Ivan Hubeny\altaffilmark{1}}
\affil{Laboratory for Astronomy and Solar Physics, \\
NASA Goddard Space Flight Center, Greenbelt, Maryland 20771}
\altaffiltext{1}{AURA/NOAO}

\and

\author{David L. Lambert}
\affil{McDonald Observatory and Department of Astronomy, University of Texas,
Austin, TX 78712}

\begin{abstract}

We test our knowledge of the atomic opacity in the solar UV spectrum. 
Using the atomic data compiled in Paper I from modern, publicly available, 
databases, we perform  calculations that are confronted with 
space-based  observations of the Sun. At wavelengths longer
than about 2600 \AA, LTE modeling can reproduce quite closely the observed
fluxes; uncertainties in the atomic 
line data account fully for the differences between
calculated and observed fluxes. At shorter wavelengths, departures from LTE 
appear to be important, as our LTE and restricted NLTE calculations differ.
Analysis of visible-near infrared Na I and O I lines, two species that
 produce a negligible absorption in the UV, shows
that observed 
departures from LTE for theses species can be reproduced very accurately
with restricted (fixed atmospheric structure) NLTE calculations. 

 
\end{abstract}

\section{Introduction}

Understanding the solar UV spectrum and its variability 
has broad multi-disciplinary interest. Firstly, it is closely connected
to life on Earth. UV radiation damages DNA, can 
cause mutations, and may affect the Earth's climate (Coohill 1996; 
Larkin, Haigh \& Djavidnia 2000).
The UV spectrum of late-type stars also has extraordinary repercussions for
the study of
larger astronomical objects.  Late-type stars have very long lifetimes, which
makes it possible to use their photospheric abundances 
 to trace the chemical evolution of the Milky Way.
Many heavy elements,  
 such as Ge, Au, Ir, Pt, or Pb, 
do not produce detectable
 features in the optical window, but 
 have strong  lines in the UV  (see, e.g., Sneden et al. 2000). 
 The same problem afflicts the 
 light elements B and Be (see, e.g., Garc\'{\i}a L\'opez 1996). 
 In an extragalactic context, understanding the formation of the UV spectrum
 of late-type stars is important because
late-type turn-off stars dominate the rest-frame mid-UV spectrum of
intermediate-age stellar populations. 
Dating an unresolved stellar system becomes then equivalent
to determining the mean effective temperature of the turn-off stars
(Heap et al. 1998). Interestingly, for intermediate and high
redshifts, the rest-frame UV is observable from Earth in the optical
and near-IR.
The age of a galaxy determined from a spectroscopic analysis 
can be in serious error if
the formation of the near-UV spectrum of late-type stars 
is not understood properly.

As astronomers realized the important role of near-UV stellar fluxes in
so many different astrophysical contexts,
a controversy arose about how well classical model atmospheres and  
available atomic data are able to reproduce observations.
Inconsistencies between observed and computed solar fluxes  were early 
reported by   Houtgast \& Namba (1968), 
Labs \& Neckel (1968),  Matsushima (1968), or
 Chmielewski, Brault, \& M\"uller (1975).  Missing line opacity  was
commonly  suggested as the origin of the discrepancy 
(Holweger 1970; Vernazza, Avrett \& Loeser 1976).  
Dragon \& Mutschlecner (1980), Kurucz \& Avrett (1981), and Kurucz (1992)
included millions of atomic and molecular lines previously ignored in
the computations, and claimed to have solved the problem. However,
Bell, Paltoglou \& Tripicco (1994) criticized the suggested solution,
as high-dispersion solar spectra in the regions 3400--3450  and
4600--4650 \AA\ revealed that the synthesis based on Kurucz's line
list  predicted many absorption lines that were not observed. 
Bell et al. (1994) and later Balachandran \& Bell (1998) and Bell, 
Balachandran, \& Bautista (2001) proposed missing
contributors to the continuum absorption (in particular, a larger Fe I
photoionization  cross-section), rather than line blanketing,
as the most likely explanation.  

The controversy continues, and the existence
of a missing UV opacity has been  
also discussed in many stellar studies. 
Malagnini et al. (1992) and Morossi et al. (1993) compared observations 
and Kurucz's calculations for late-G and early-K stars, finding that theory
underpredicted the observed near-UV fluxes. Other
authors have apparently not found such inconsistencies in the analyses
of late-type metal-poor stars, or  early-type stars 
(Fitzpatrick \& Massa 1999; Allende Prieto \& Lambert 2000; 
Peterson, Dorman, \& Rood 2001).

In Allende Prieto et al. (2002, Paper I), we presented a new set of 
model atoms for use in 
NLTE calculations  that are  based on harvesting modern databases 
following a set 
of uniform criteria.  We use the new model atoms here to
study  the sources of opacity 
in the near-ultraviolet
spectrum of the Sun, and take a fresh look at the problem of the
missing opacity (\S \ref{uv}).
We apply the data to the so-called restricted NLTE problem -- the solution
of the statistical equilibrium equations adopting a fixed LTE structure --
for the solar photosphere, examining the effect of departures from LTE in
optical and near-infrared oxygen and sodium 
line profiles in \S \ref{lines}.  Section 4 summarizes
our results.

\section{The solar near-UV continuum}
\label{uv}

To produce a realistic calculation of the solar UV flux is vastly more
complicated than doing so at visible or near-infrared wavelengths. 
The penetration of convection
into the photosphere results in thermal and velocity inhomogeneities
(granulation), 
whose effect on the spectral energy distribution has never been seriously 
considered. 
At wavelengths shorter than 2500 \AA, bound-free 
metal absorption becomes dominant,  
producing a sharp increase in opacity  towards shorter wavelengths, as we cross
the different metal photoionization thresholds.
This is a serious difficulty, as  photoionization 
cross-sections for metals are not as well determined as for H or H$^{-}$. Line
absorption becomes
 a very important contributor at shorter wavelengths, 
 introducing an additional
 obstacle to accurate modeling because of the limited availability
 and quality of the required  atomic
data. Finally, the increase in  opacity shifts 
the formation region to higher atmospheric layers, where the density is low, 
 and spatial (and time) inhomogeneities appear to be significant. 
 This last difficulty may be 
so serious that most of the assumptions adopted in  photospheric models 
could break down.

Nonetheless, much progress has
occured in the last several years affecting availability of atomic data,  
observations, and modeling techniques, and thus it seems appropriate to
use the new model atoms to update the calculations and confront
them  with observations. We adopt a {\it fixed} atmospheric structure
(temperature, electron pressure and gas pressure
 as a function of optical depth)
calculated assuming LTE, energy and hydrostatic equilibrium, and a plane-parallel
geometry.  
We employ the necessary atomic data (Paper I) 
 together with the
 hybrid Complete linearization/Accelerated Lambda
Iteration method (Hubeny \& Lanz 1995) to solve 
the so-called restricted NLTE problem for different species.

\subsection{Modeling}

Our analysis ignores a number of  factors in the modeling that are 
known to affect the computed fluxes. The adopted values for the mixing-length 
parameters (see, e.g., Castelli, Gratton, \& Kurucz 1997; 
Barklem et al. 2002), or the adopted solar abundances are
two controversial examples. In addition, we neglect molecular opacities, which
are known to contribute most at wavelengths shorter 
than about 2200 \AA.
To keep the number of variables at a
manageable level, concentrating on the effect of the metal-opacities, we
have chosen to use an LTE 
model atmosphere  
interpolated from the grid of Kurucz (1993). The employed grid was
calculated for  a microturbulence of 2 km s$^{-1}$, and a mixing-length
$\alpha=l/H_p = 1.25$ with no overshooting. The adopted model
is indeed quite similar to the {\tt MISS} 
semi-empirical model of Allende Prieto et al. 
(2001a; see Fig. \ref{fig1}), and the calculated fluxes are 
very close in the wavelengths we are
interested in (see below), but the theoretical 
structure a) extends into smaller optical depths, and b) is 
originally specified in terms
of the column density (mass in Fig. \ref{fig1}), rather than optical depths, and 
therefore we avoid the use of external opacity packages to convert between 
those two scales. We have adopted a value for the microturbulence of 1.1 
km s$^{-1}$ (Allende Prieto et al. 2001a), and solar abundances from
Grevesse \& Sauval (1998), unless otherwise specified. 

\begin{figure*}
\begin{center}
\includegraphics[width=10cm,angle=90]{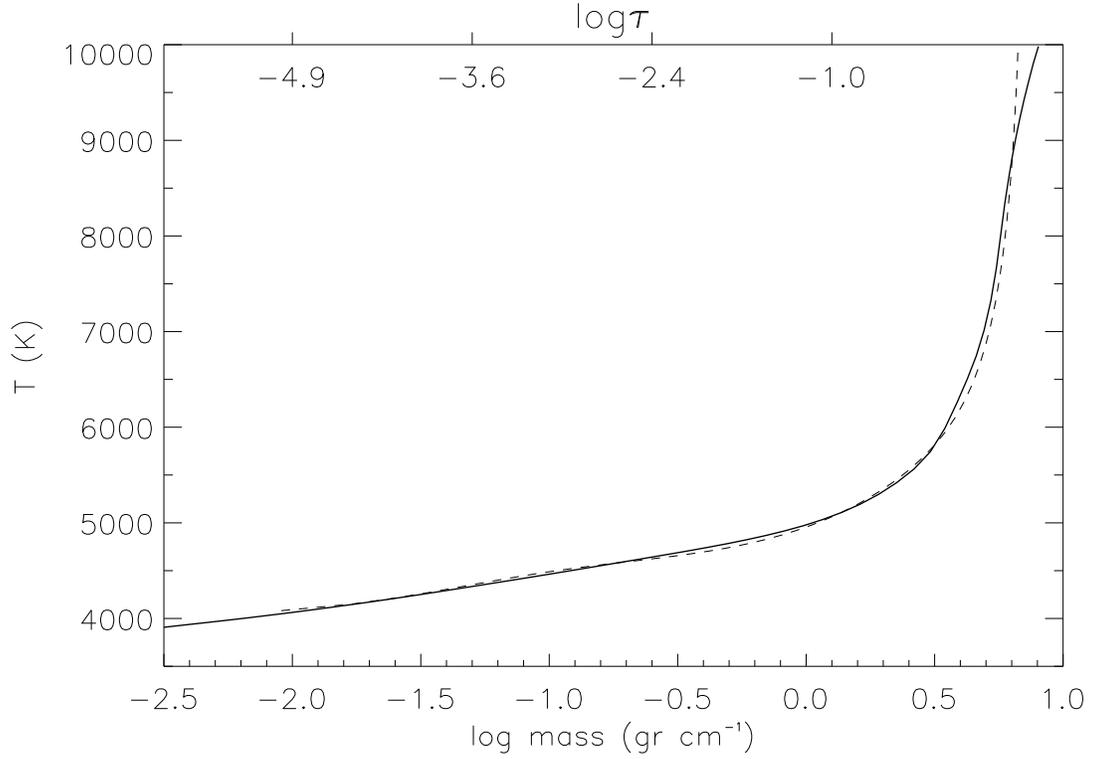}
\figcaption{
Comparison between the theoretical model atmosphere used in our calculations
(solid line)
with the {\tt MISS} semi-empirical model derived 
by Allende Prieto et al. (2001; dashed). 
\label{fig1}}
\end{center}
\end{figure*}

We considered the following 
species in NLTE: H I, He I, Li I, Be I, Be II, B I,
C I, N I, O I, Na I, Mg I, Mg II, Al I, Si I, Si II, Ca I, Ca II, Fe I, and
Fe II.  Our model atoms for Fe I and Fe II are based on the concept of
superlevels, described in Hubeny \& Lanz (1995; see also
Anderson 1989). We considered H$^{-}$ opacity,
as well as electron scattering and Rayleigh scattering by neutral hydrogen
(see Hubeny 1988 for more details).
Besides the largely dominant H$^{-}$, we found C I, 
Mg I, Al I, Si I, Ca I, and Fe I to be relevant for the near-UV continuum of 
a solar-type star.

\begin{figure*}
\begin{center}
\includegraphics[width=10cm,angle=90]{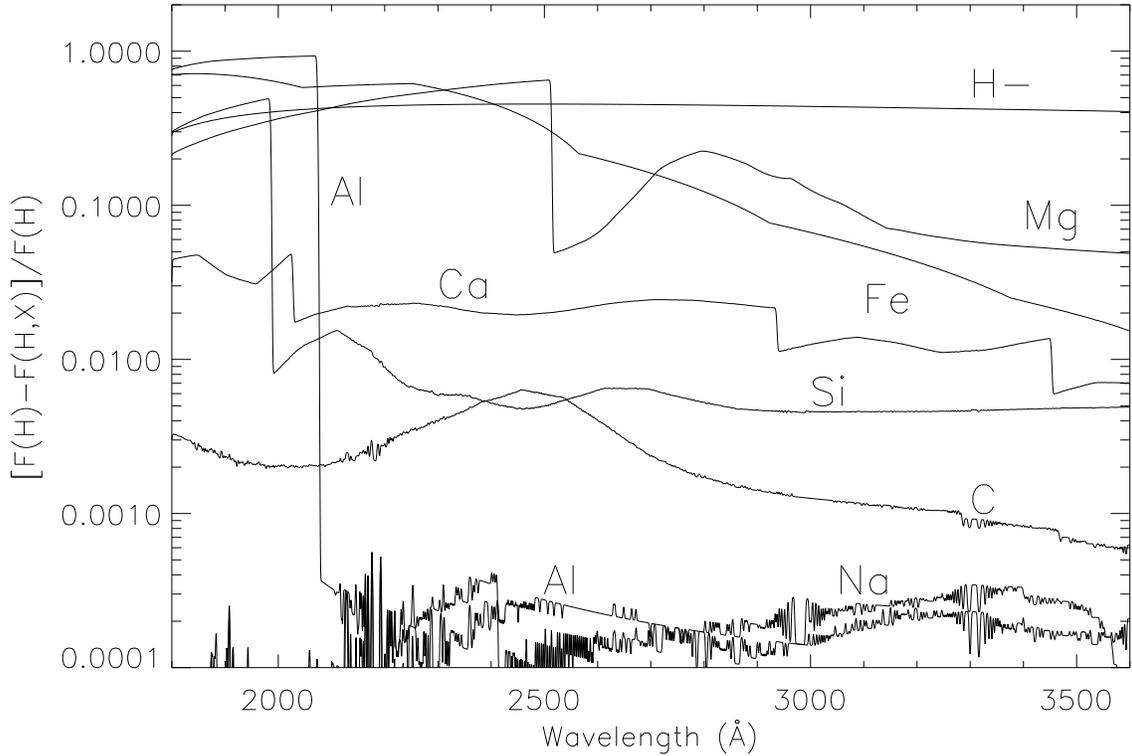}
\figcaption{
Relative difference between the continuum (bound-free and free-free
opacity only) flux when only H opacity is considered and when a second 
contributor is introduced. The different curves reflect the 
contribution to the continuum absorption relative to the H opacity 
in the solar UV spectrum made by the following species and elements: 
H$^{-}$, C, N, Na, Mg, Al, Si, Ca, and Fe. Round-off errors in our 
calculations become apparent for relative differences under $\sim 10^{-3}$.
\label{fig2}}
\end{center}
\end{figure*}

\begin{figure*}
\begin{center}
\includegraphics[width=8.2cm,angle=0]{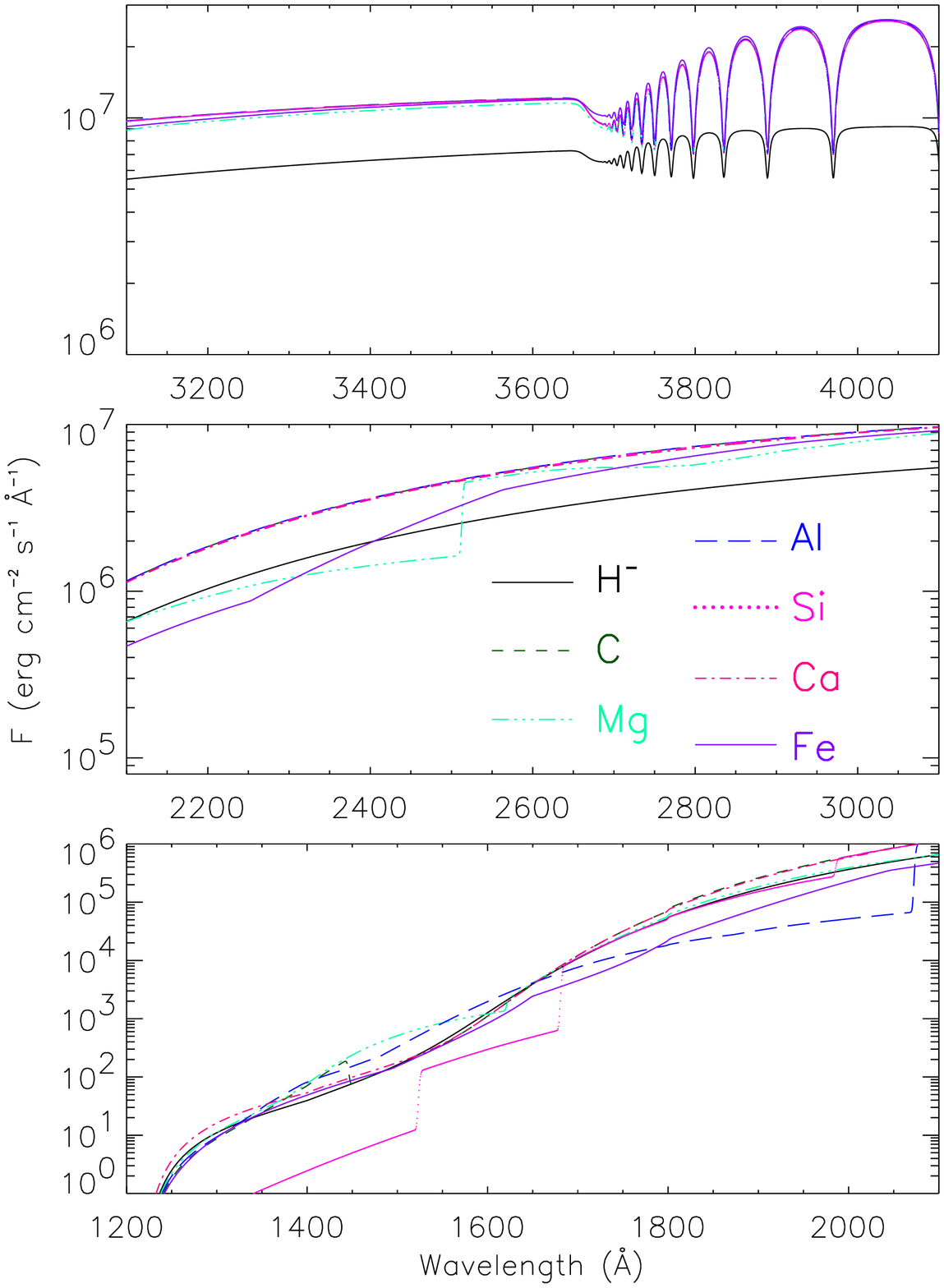}
\includegraphics[width=8.2cm,angle=0]{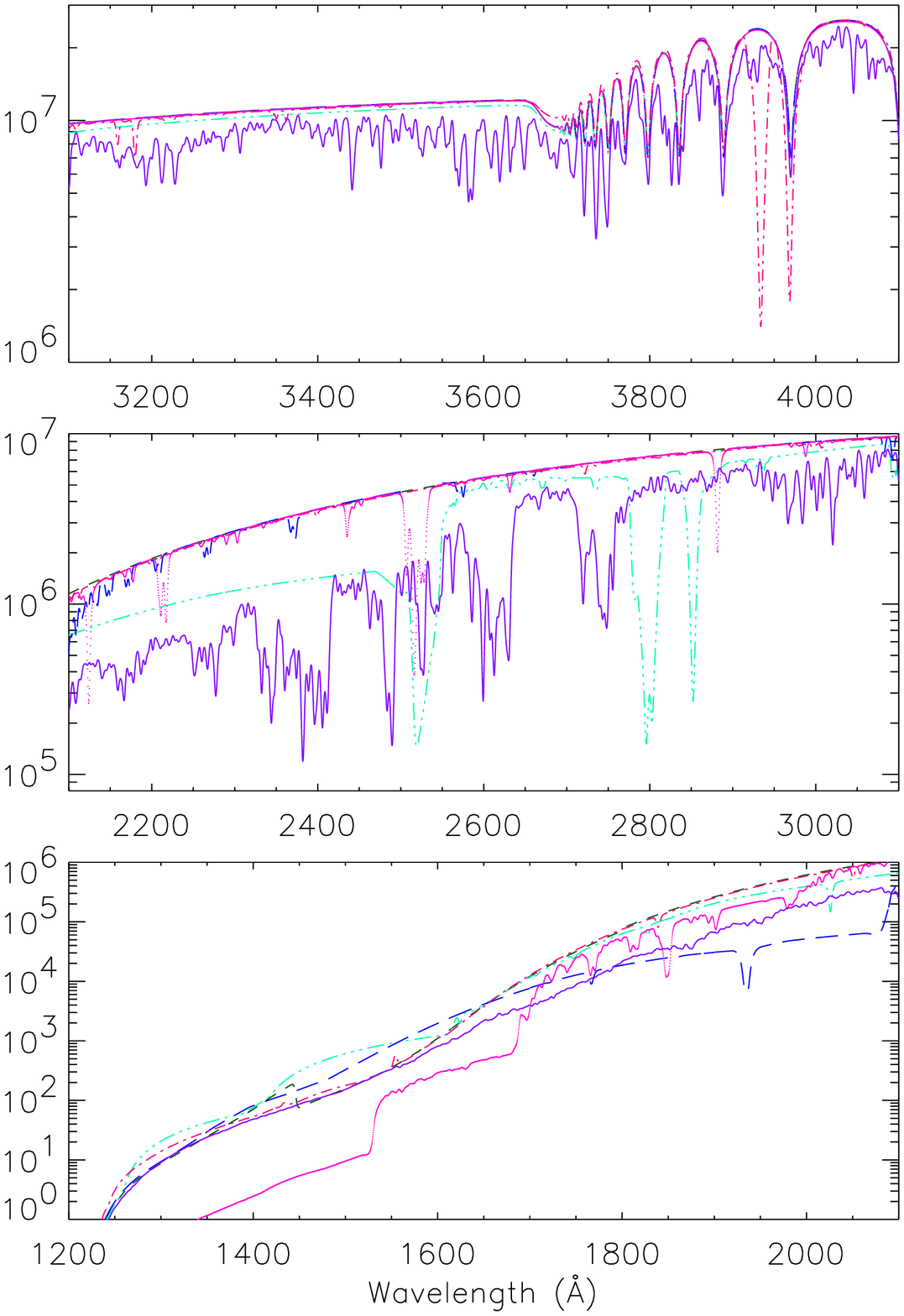}
\figcaption{
{\it Left:} Emergent flux when only 
the continuum absorption  produced by H $^{-}$, C, N, Na, Mg,
Al, Si, Ca, and Fe is considered; {\it Right:} Emergent flux when the
 total (continuum + line) opacity  is considered. The curve corresponding to
 H$^{-}$ has been omitted in the right-hand panels for clarity.
\label{fig3}}
\end{center}
\end{figure*}

Fig. \ref{fig2} shows the normalized difference between the predicted LTE fluxes 
when only hydrogen (bound-free, free-free and
bound-bound) opacity is included, F(H), and  the only addition of one more
contributor, F(H,X), where X can be H$^{-}$, or any of the metals. 
The graph shows the impact of the different 
species, in such a way that the closer to unity they reach
in the plot,
the more important their contribution is. The contribution from 
the metals comes from the bound-free 
absorption by neutral atoms, mainly from the lowest levels.
This comparison 
shows that besides neutral hydrogen and H$^{-}$, only Fe I and Mg I
are relevant at wavelengths longer than 2100 \AA. Al I becomes important
for wavelengths shorter than  about 2070 \AA. Below that wavelength the figure
is not so informative, as H loses its dominant role. 
Fig. \ref{fig3} shows the actual LTE 
fluxes F(H,X) excluding (left panel), and including (right panel), 
line opacity, revealing that  Si I 
bound-free absorption  becomes significant 
for wavelengths shorter than  about 2000 \AA, and dominant below $\sim 1670$
\AA. As discussed below, the photospheric origin of the  solar 
spectrum observed at these and shorter wavelengths is questionable, and
the adequacy of the adopted model atmospheres more than 
doubtful. Our  results are in qualitative agreement with previous studies 
\footnote{Comparison
with previous works shows that Cr I will also produce a very small 
(but noticeable in Fig. \ref{fig2}) contribution to the near-UV continuum 
opacity.} (e.g. Travis \& Matsushima 1968, Bell et al. 2001).

\subsection{Observations}

In the last decade two instruments aboard the 
Upper Atmosphere Research Satellite (UARS) have
provided solar near-UV absolute irradiances with
unprecedented accuracy. The Solar-Stellar Irradiance Comparison Experiment
(SOLSTICE; Rottman, Woods, \& Sparn 1993; 
Woods, Rottman, \& Ucker 1993)  has 
continuously monitored the solar spectrum between 1190 and 4200 \AA\
 since 1991 with a resolution of up to 2.5 \AA. 
Flying on the same spacecraft, the Solar Ultraviolet Spectral Irradiance 
Monitor (SUSIM; Brueckner et al. 1993) 
has a similar spectral coverage and resolution. Both 
instruments had pre-flight calibrations using the Synchrotron Ultraviolet
Radiation Facility (SURF II) at the National Institute of Standards and
Technology (NIST), but SOLSTICE observes hot stars to track variations in the 
instrumental sensitivity, whereas SUSIM uses standard lamps and redundant
optics for that task. 

Additional checks of the absolute solar fluxes measured by these 
instruments were carried out by comparison to simultaneous measurements 
with two more instruments, a copy of SUSIM-UARS (SUSIM ATLAS) and the Shuttle
Backscatter UltraViolet experiment (SSBUV),
that flew on the shuttle ATmospheric Laboratory
for Applications and Science (ATLAS) missions in March 1992, April 1993, and
November 1994 (Woods et al. 1996). 
SUSIM-UARS daily solar irradiances (version 20; Level 3BS) 
between 1991 and 1998 are  available from the 
web\footnote{http://louis14.nrl.navy.mil\/susim\_uars.html}. SOLSTICE
daily averaged spectra (version 9; Level 3BS) between 1991 and 1996 are also publicly available
\footnote{ftp://daac.gsfc.nasa.gov/data/uars/solstice/}.  

\begin{figure*}
\begin{center}
\includegraphics[width=7cm,angle=90]{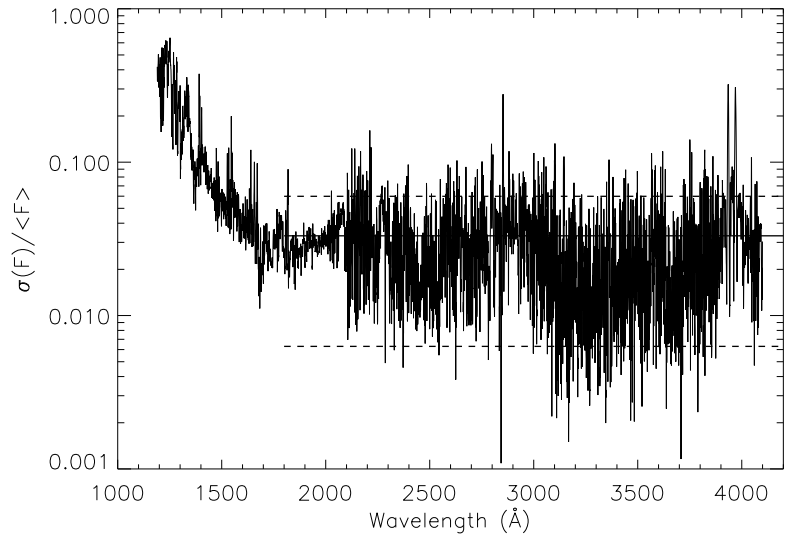}
\figcaption{
Relative standard deviation of six solar spectra acquired by SOLSTICE and
UARS/SUSIM in the different epochs of the ATLAS flights: March 1992, April 1993,
and November 1994. The solid line represents the mean value for 
$\sigma({\rm F})/<{\rm F}>$, and the dashed lines 
show the mean $\pm 1 \sigma$ for $\lambda > 1800$ \AA. Real solar chromospheric
variability is likely dominant at shorter wavelengths.
\label{fig4}}
\end{center}
\end{figure*}

We have combined six spectra obtained by SOLSTICE and SUSIM-UARS during the
three flights of the ATLAS mission. The SUSIM spectra were smoothed to the same
resolution as the SOLSTICE data by T. Woods, and are available from the SOLSTICE
web site\footnote{http://lasp.colorado.edu/solstice/}. 
The standard deviation of the six, divided by their 
mean is shown in Figure \ref{fig4}. These figures provide an 
order-of-magnitude estimate of the uncertainty
in the calibrations and, at short wavelengths,  
to the level of variability of the solar spectrum in the period bracketed by
the observations. 
For wavelengths longer than about 1800 \AA, we can safely compare our 
calculations to the average spectrum within a few percent. The UARS fluxes
are given as irradiance at Earth's mean distance (1 AU).

The uncertainties in the adopted value for the
astronomical unit, the corrections between the position of the observatory
and the center of the Earth, and the displacement of the solar center from
the Earth-Sun barycenter,  are small 
enough to be neglected (Huang et al.
1995; Brown \& Christensen-Dalsgaard 1998). The correction of the observed
irradiance to determine 
the flux at the solar surface deserves some comments. A comparison
among published measurements of the apparent solar radius 
between 1980 and 2000 by Golbasi et al.
(2001) shows a large scatter. It is unclear whether the listed
measurements can be directly compared, and  which of the employed 
methods (and definitions) is the more appropriate to use in our case. 
The many contradictory results in the literature reporting variations of the 
solar radius with the solar cycle, and on other time scales, alerts us to 
systematic effects between different experiments (see, e.g., Basu 1998;
Antia et al. 2000).

Five average measurements (from three different methods) 
centered in the years between 1990
and 1995 show relatively consistent results, with an average value
$\theta/2=959.492 \pm 0.085$ arcsec. At the same time, the data suggest
a slow decrease in $\theta$ with time, which seems also in agreement with
the results of Costa et al. (1999) from radio maps. 
Conservatively, we adopt an uncertainty for 
 $\theta/2$ of $0.25$ arcsec, and therefore we
 correct the observed flux to 
determine the solar surface flux by applying a factor 
(d/R)$^2 = 1/\tan^2(\theta/2) =
46250 \pm 40$. This factor introduces less than a 0.001 \% uncertainty in the
derived solar surface flux, 
which is negligible compared to the scatter among the
six SOLSTICE/SUSIM spectra. If we use the solar radius derived from the 
observed p-mode frequencies by Takata \& Gough (2001; see also the
results from the f-modes by Schou et al.
1997 and Antia 1998), we find (d/R)$^2=  46240 \pm 19$. A decrease  
in the total solar irradiance by $\sim 0.02$ \% 
between November 1991 and 1995 has  been also
reported from measurements from 
another instrument on UARS,  ACRIM II (Fr\" ohlich 2000). 
This reduction in the solar irradiance is about four times larger than what
could be induced by the possible shrinking in $\theta$.
If interpreted as a change in the solar effective temperature, such variation 
would be produced by a decrement in $T_{\rm eff}$ of less than two degrees.

\subsection{Observations vs. calculations}

Some of our calculations including line opacity employ the line list prepared 
by Kurucz and distributed with 
TLUSTY\footnote{http://tlusty.gsfc.nasa.gov}. 
A newer line list is available from 
Kurucz's web site\footnote{http://kurucz.harvard.edu}. Further tests were
carried out with line list based on a {\sc stellar} request to the Vienna
Atomic Line Database (VALD; Kupka et al. 1999; Stempels, Piskunov, \&
Barklem 2001) for the 
stellar parameters of the Sun. Although
the differences were small, we obtained a marginally better agreement with
the observations in the 2700--4000 \AA\ region when using the
VALD list. Therefore, the VALD list was adopted.

\begin{figure*}
\begin{center}
\includegraphics[width=12cm,angle=0]{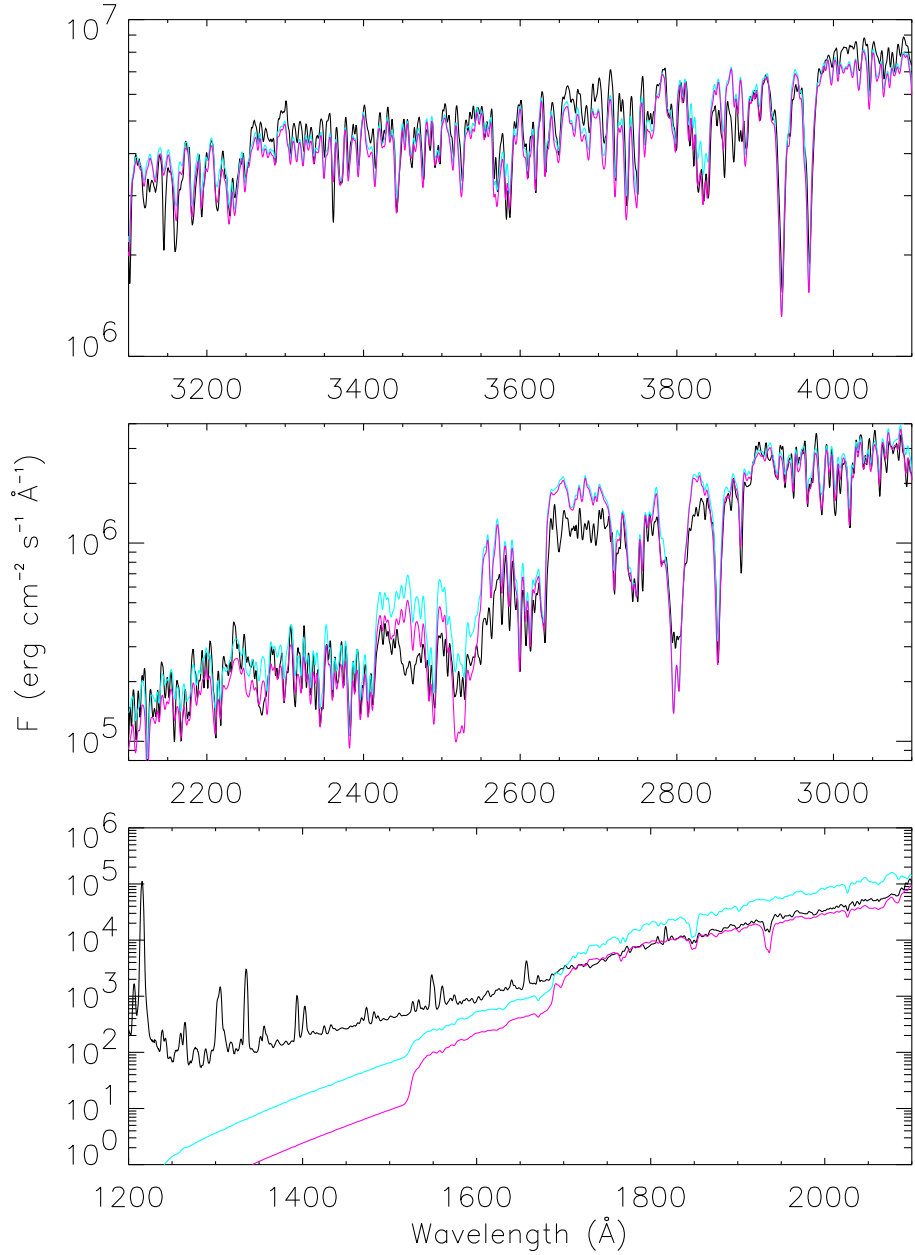}
\figcaption{
Comparison between the observed (black) and computed fluxes for the solar
UV spectrum. The calculations are always based on an LTE atmospheric 
structure, and the level populations were computed assuming LTE (red line) 
or by solving the restricted NLTE problem for the 
relevant species (blue line).
\label{fig5}}
\end{center}
\end{figure*}

Fig. \ref{fig5} shows  our LTE calculation and the observed spectra. 
The overall agreement at wavelengths longer than about 1750 \AA\ is remarkable, 
especially to the red of $\sim 2700$ \AA. With the exception of a few
isolated regions, the observed fluxes are closely reproduced. 
Between 2700 and 4000 \AA, the predicted fluxes are  an average of 1.8 \%
lower than the observations ($\sigma_{\rm rms}$= 14 \%). If we use the
{\tt MISS} semi-empirical model (Allende Prieto et al. 2001a), 
we find a slightly larger average 
difference of 7 \% ($\sigma_{\rm rms}$ = 13 \%). At these wavelengths, diatomic
molecules may introduce some additional line opacity that we have neglected,
but the good agreement that we find with the observed fluxes suggests that
this must be very limited.
The disagreement found in the shortest wavelengths is not a surprise, 
 as the large 
increase in opacity at these wavelengths 
shifts the atmospheric region where the spectrum is formed from the
photosphere to the chromosphere. 
Below $\sim 1600$ \AA\, the continuum is formed at $\tau \lesssim 10^{-4}$. 
Dragon \& Mutschlecner (1980) pointed out that at wavelengths 
shorter than about 2200 \AA\ CO line opacity becomes very important.
Our own LTE calculations including about 400,000 CO transitions with data 
 from Kurucz's line lists indicate, however, that CO is only a minor 
contributor between 1100 and 4600 \AA. 

In the top panel, where the agreement is best, a systematic discrepancy
is apparent to the red of 4000 \AA. The continuum opacity is well determined at
those wavelengths and, therefore, this might signal problems with the 
calibration of the observed fluxes in that region -- extreme for the 
discussed instruments. 
Between 1700 and 2200 \AA, the LTE calculation shows better agreement
with the observations. This is most likely connected to the 
photoionization edge
of the ground level of Al I. The Al I autoionization line at 1932 \AA\ is 
 predicted in LTE and observed, but it is missing in the NLTE calculation, 
confirming our suspicions about the NLTE population of the ground state of Al I.
Conversely, the NLTE flux is closer to the observations than the LTE 
calculation for the range 2200--2400 \AA.
In the middle panel of Fig. \ref{fig5}, the region 
between 2400 and 2600 \AA\
shows a poor fit, which, 
from inspection of Fig. \ref{fig3}, is probably connected to the
 photoionization of Mg I. The predicted departures from LTE for 
 the lowest levels of Mg I seem to have the wrong sense in that 
 our NLTE calculation   enhances the 
 discrepancy with the observations in this spectral window. Dragon \&
 Mutschlecner (1980), and Kurucz \& Avrett (1981) also noticed significant
 discrepancies between observed and computed fluxes in this region. 
At 2650--2700 \AA, the disagreement may also be caused by departures
 from LTE in  neutral magnesium or iron.

\begin{figure*}
\begin{center}
\includegraphics[width=7.5cm,angle=90]{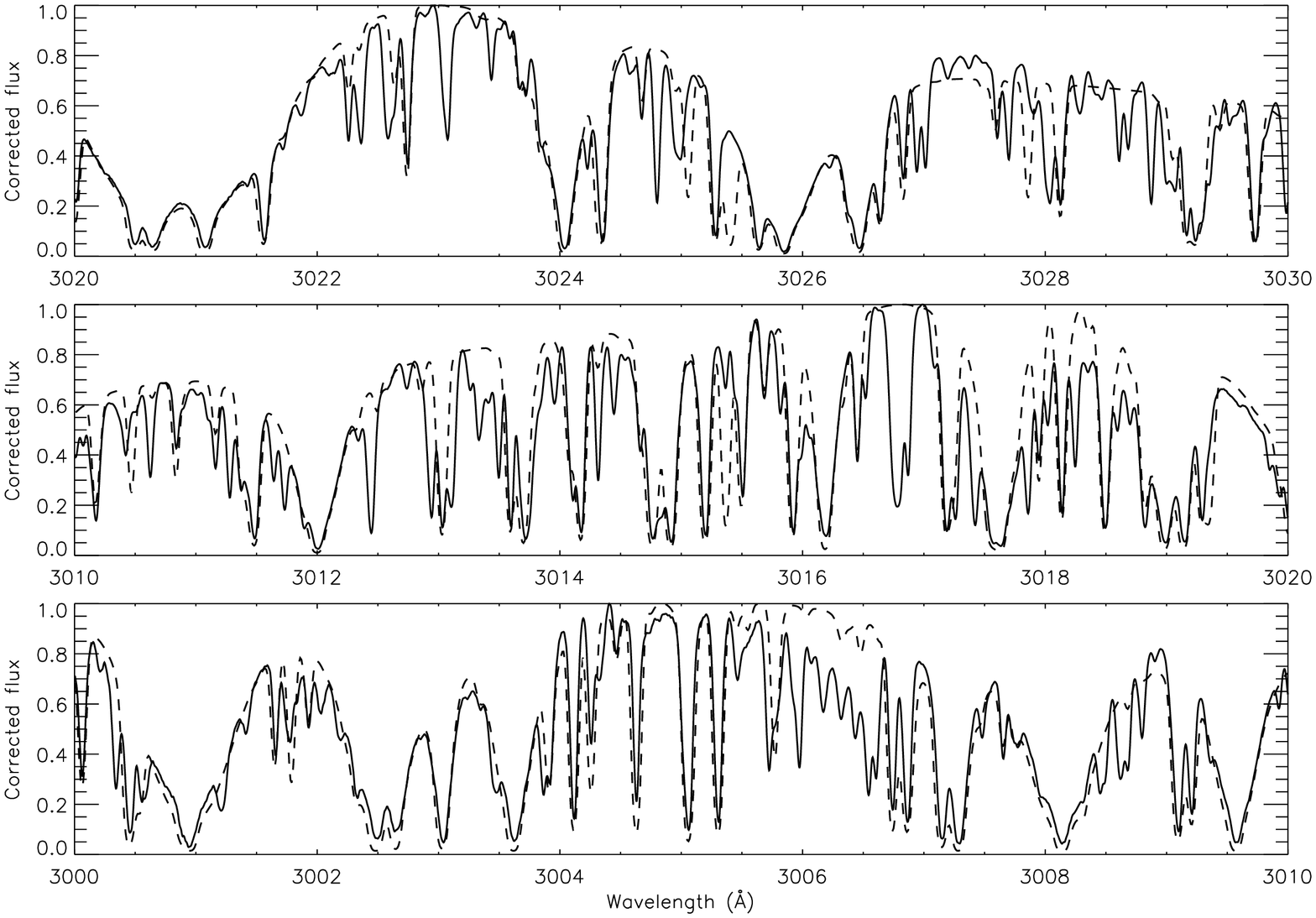}
\includegraphics[width=7.5cm,angle=90]{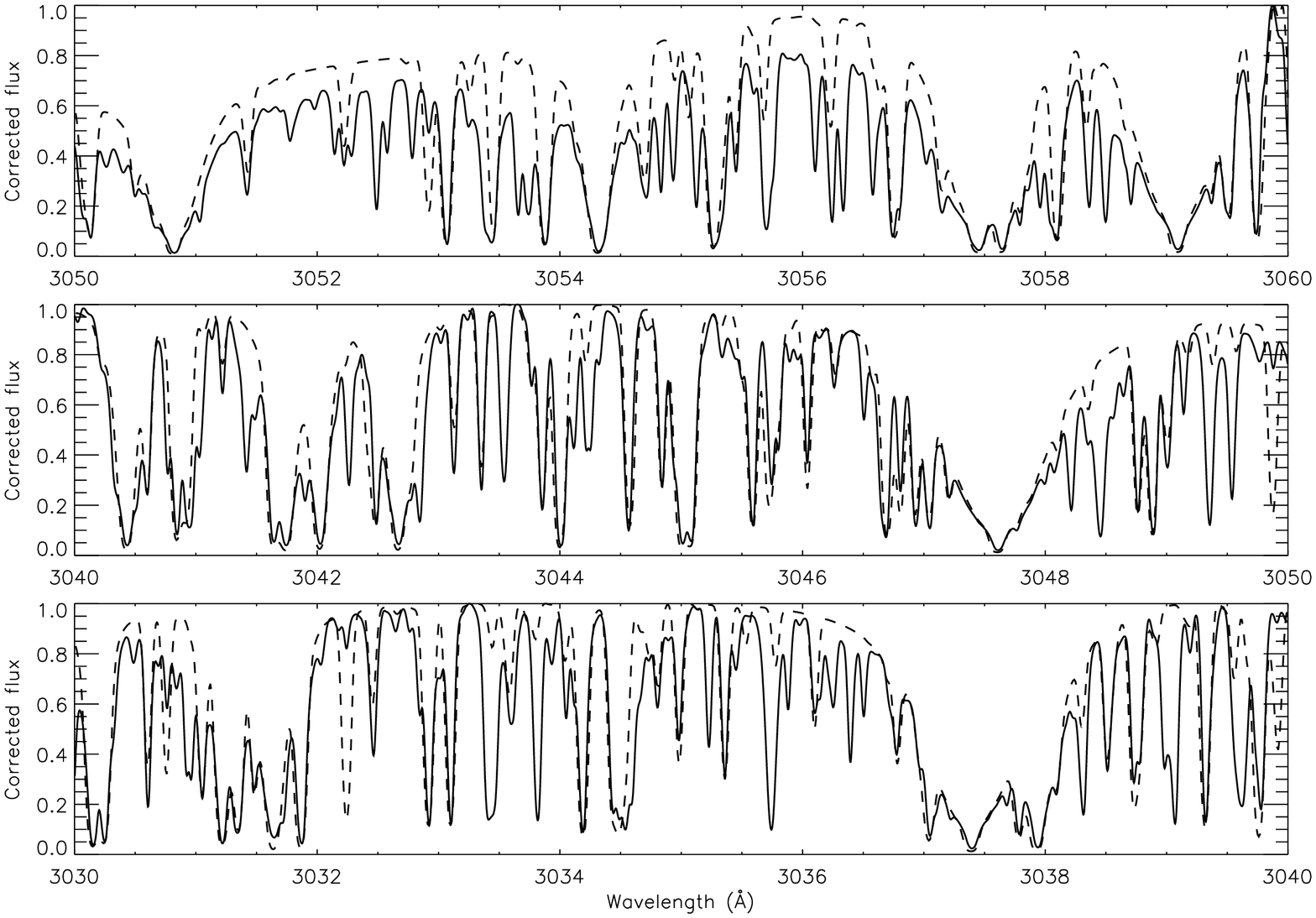}
\includegraphics[width=7.5cm,angle=90]{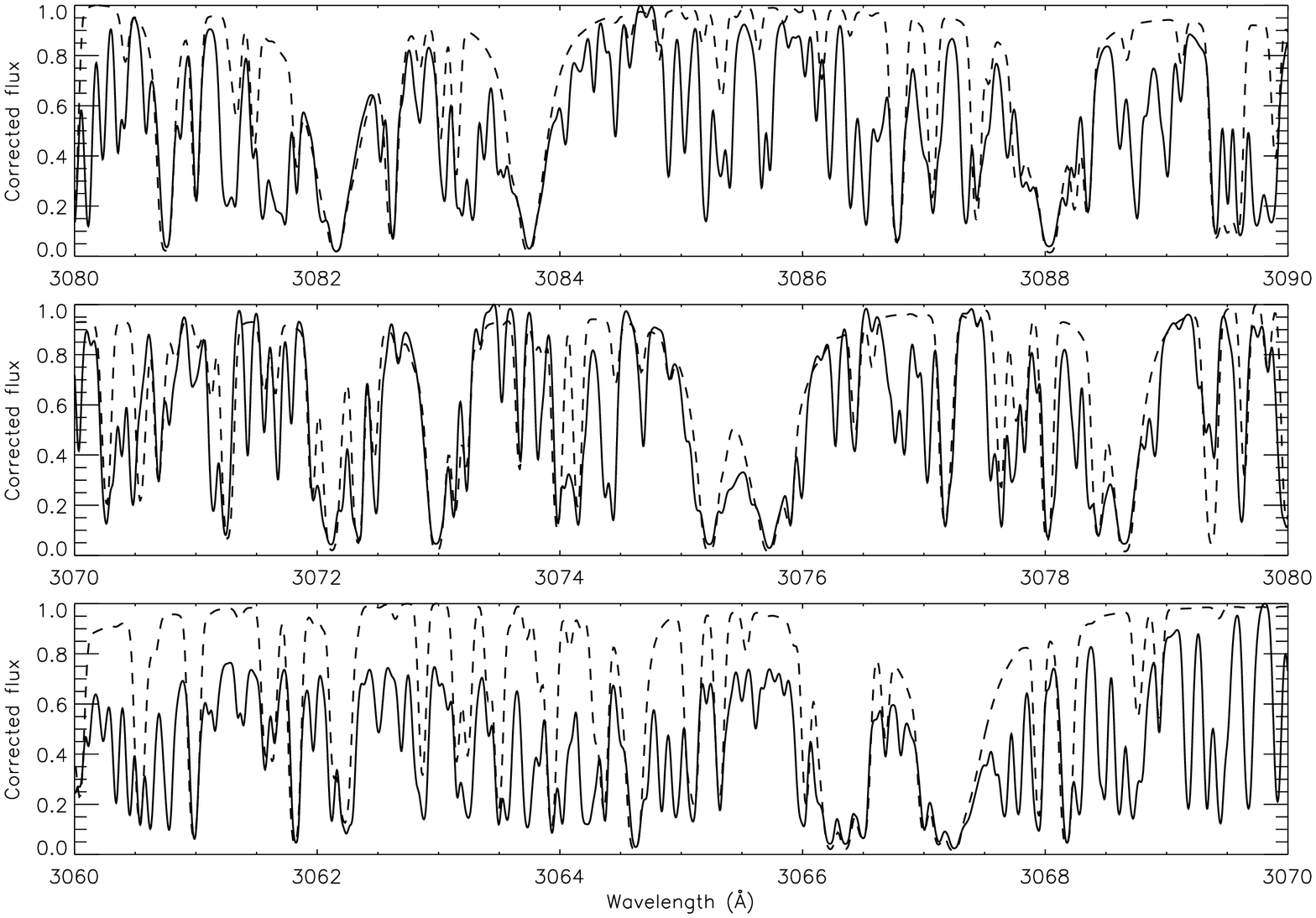}
\figcaption{
The  observed (solid line) and calculated (dashed) fluxes at very high resolution in the range
3000--3090 \AA\ are fit to and divided by a straight line derived by
least-squares fitting to compare the calculated and real line absorption. 
The maximum flux has been normalized to 1 for each window. OH lines are not
included in the line list used in the spectral synthesis.
\label{fig6}}
\end{center}
\end{figure*}

\begin{figure*}
\begin{center}
\includegraphics[width=10cm,angle=90]{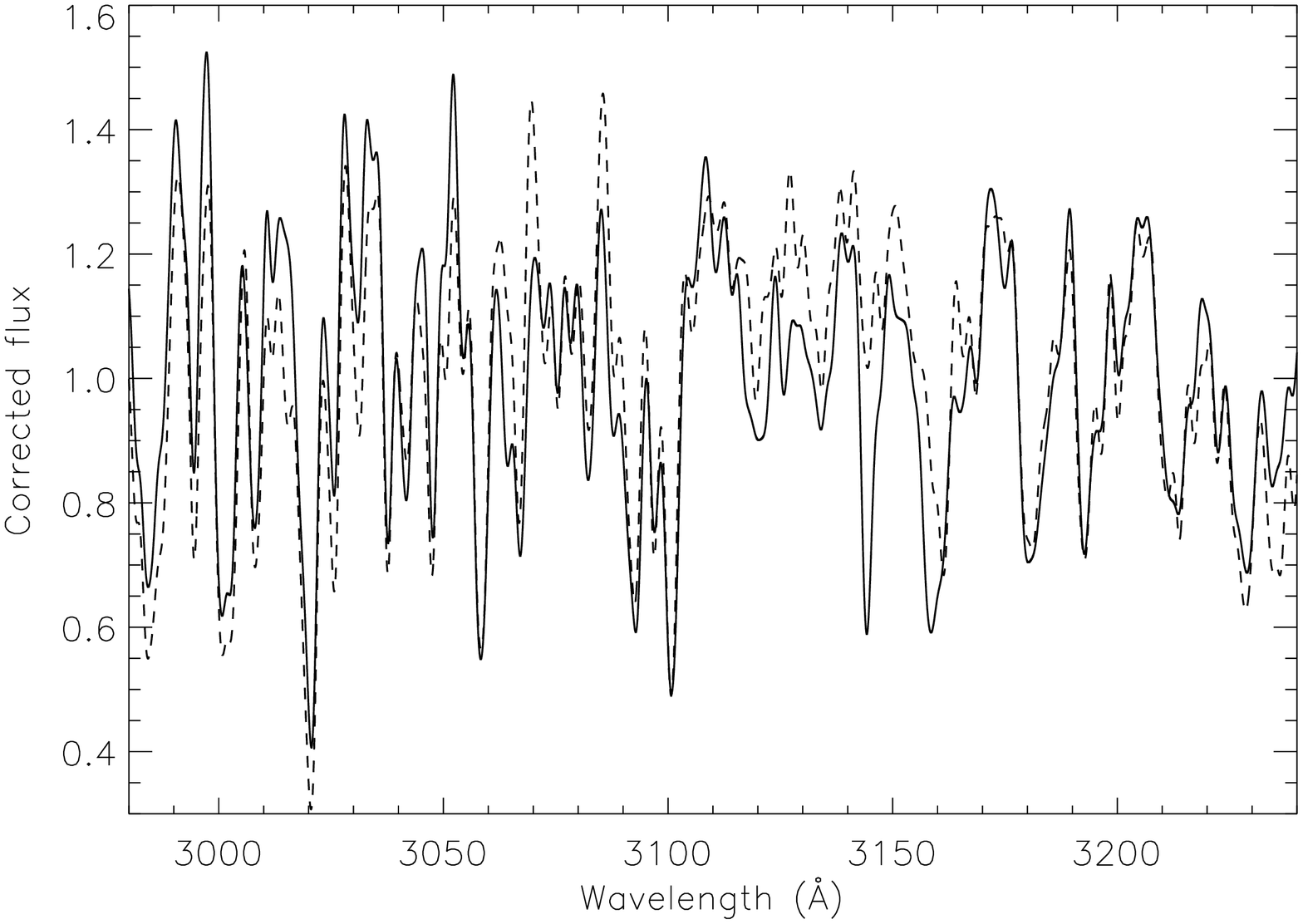}
\figcaption{
The  observed (solid line) and calculated (dashed) fluxes at very high resolution in the range
2980--3240 \AA\ are fit to and divided by a straight line derived by
least-squares fitting to compare the calculated and real line absorption. The
comparison shows the fluxes after convolution with a Gaussian with FWHM 2.0
\AA, to remove the high-frequency variations and keep those we are
interested in for comparing with the SOLSTICE/SUSIM observations.
\label{fig7}}
\end{center}
\end{figure*}

It is possible to compare the LTE calculations at very high resolution  
($R= 5 \times 10^5$) with the solar atlas of Kurucz et al.  (1984) 
for wavelengths redwards of 2960 \AA. 
Fig. \ref{fig6} compares the high-resolution observations in the solar
atlas of Kurucz et  al. (1984) with the synthesis at a similar 
resolution between 3000 and 3090 \AA.
Since the observed spectrum is not in an absolute flux scale, 
we have normalized the two datasets maxima to be 1 in each window.
A comparison at lower dispersion is even more relevant for our goals.
We have adjusted the observed and calculated spectra 
in the window 2980--3240 \AA\ by fitting to, and dividing by, 
a straight line each of them.
Fig. \ref{fig7} shows the corrected fluxes after degradation 
to a resolution of 2 \AA. 
These Figures indicate that excessive line absorption is not 
covering  a lack of continuum opacity. 
The $\sigma_{\rm rms}$ between the two curves in Fig. \ref{fig7} is 13 \%.
This leads us to suggest that uncertainties in the line data 
dominate the random errors in the synthetic fluxes, and that the observed 
absolute fluxes are well reproduced by our calculations, at least for
$\lambda > 2700$ \AA.

\section{The restricted NLTE problem in the solar atmosphere. Optical and
near-infrared line profiles of O I and Na I.}
\label{lines}

From our comparison with observed fluxes we can expect that  using
a fixed LTE atmospheric structure to solve the level populations of
species with a important role at short ($\le 2000$ \AA) wavelengths will
be problematic, as the continuum flux is not well matched in that region.
We carried out such calculations for Mg I and Al I, finding 
that the observed line profiles of optical transitions were indeed generally  
very poorly matched in NLTE, or at least the NLTE predicted profiles differed
more significantly  from the observed ones than did the LTE profiles.
Our statistical equilibrium calculations rely on the concept of superlevels
to account for the iron line absorption. Explicit consideration of the 
line absorption -- line by line, as we do for the spectral synthesis in
Fig. \ref{fig5}-- may be necessary to obtain realistic NLTE populations for
species like Al I and Mg I.

An independent check of the atomic models can be 
performed for the species that do not produce significant
absorption in the UV. We have selected two elements with particular
importance in stellar astrophysics: oxygen and sodium. These two elements 
are very weakly sensitive to the UV flux, and they have essentially 
no effect on
it (see \S 2). 
All the calculations were carried out with the model atoms
described in Paper I.

We aim only  at identifying 
  key aspects of the NLTE modeling and, therefore, following 
  Allende Prieto et al. (2001a), we invariably adopt a depth-independent 
  micro-turbulence of 1.1 km s$^{-1}$, and a Gaussian macro-turbulence of 
  1.54 km s$^{-1}$. We also remind the reader
  the neglect of fine structure in the model atoms -- which are likely
responsible for small differences between lines of the same multiplet. The
  spectral syntheses made use of the damping constants for H collisions 
  published by Barklem, Piskunov, \& O'Mara (2000), assuming the cross-section
  to be proportional to T$^{0.4}$ for all lines. The $f-$values for the 
  discussed
  lines are well known -- as indicated
   by the quality flag in the NIST database --
  or adopted from previous works in the literature that are explicitly
  mentioned in the corresponding sections.

\subsection{Oxygen}

The O I triplet at about 7770 \AA\ can be observed in quite different 
types of stars, and it is the only strong atomic feature produced by oxygen in 
late-type stellar spectra. Despite the high excitation of this multiplet, 
NLTE effects are important (e.g. Kiselman 1993, 2001).  
A recent analysis of the [O I] forbidden line
at about 6300 \AA, which seems unaffected by departures from LTE, 
provides a reference abundance which, in addition to the line profiles, may
help to validate our NLTE calculations.

There is a controversy
as to the importance of hydrogen collisions 
 in the statistical equilibrium of oxygen 
 (e.g., Reetz 1999; Takeda 1994).   We carried
  out some tests with  
  different model atmospheres finding small, but appreciable 
  changes in the predicted lines. We take into account the charge transfer
  reaction: O + H$^{+}$ $\rightarrow$ O$^{+}$ + H, with  the data compiled by 
  Kingdon \& Ferland (1996).

\begin{figure*}[top!]
\begin{center}
\includegraphics[width=9cm,angle=0]{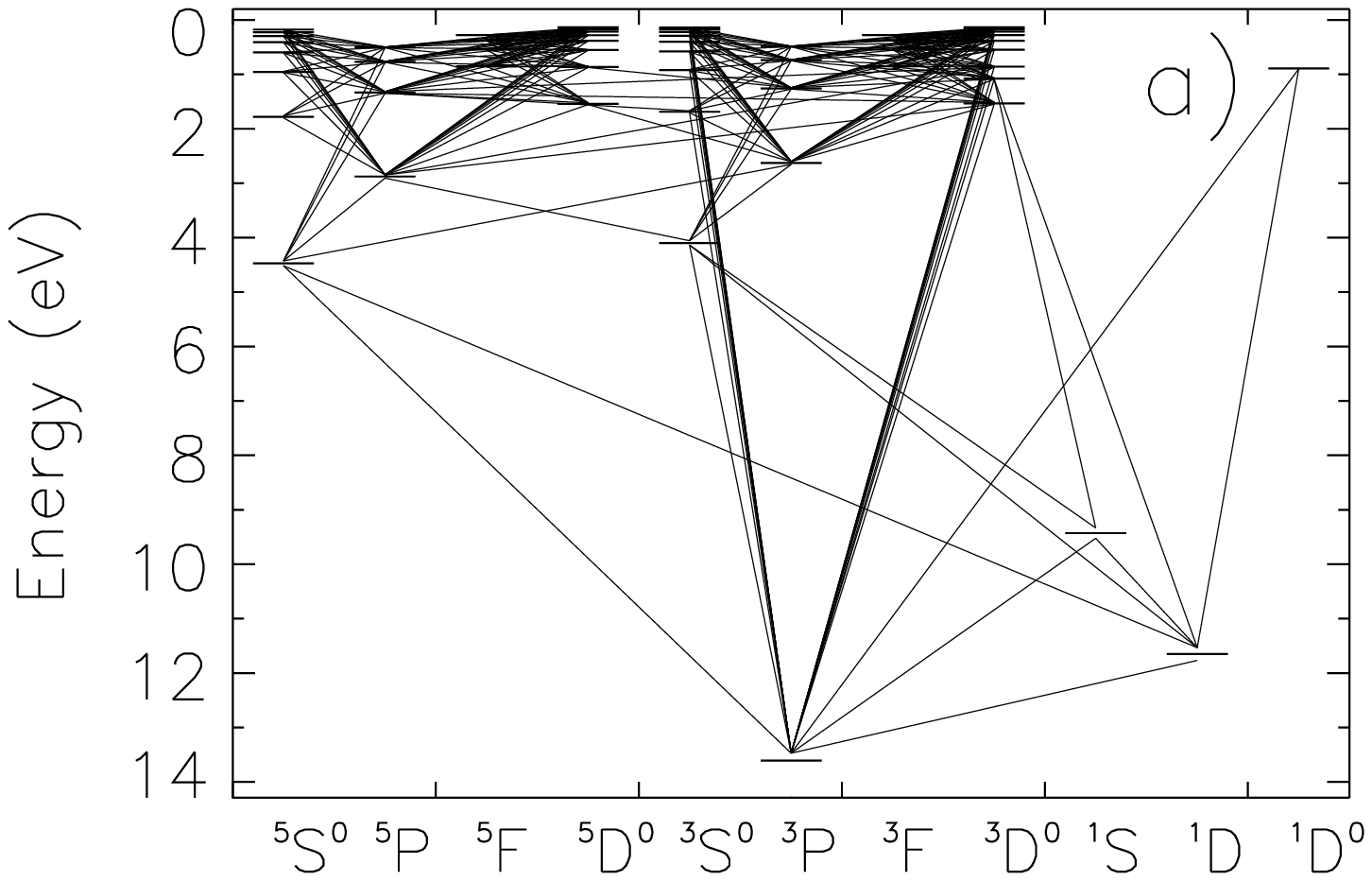}
\includegraphics[width=8cm,angle=0]{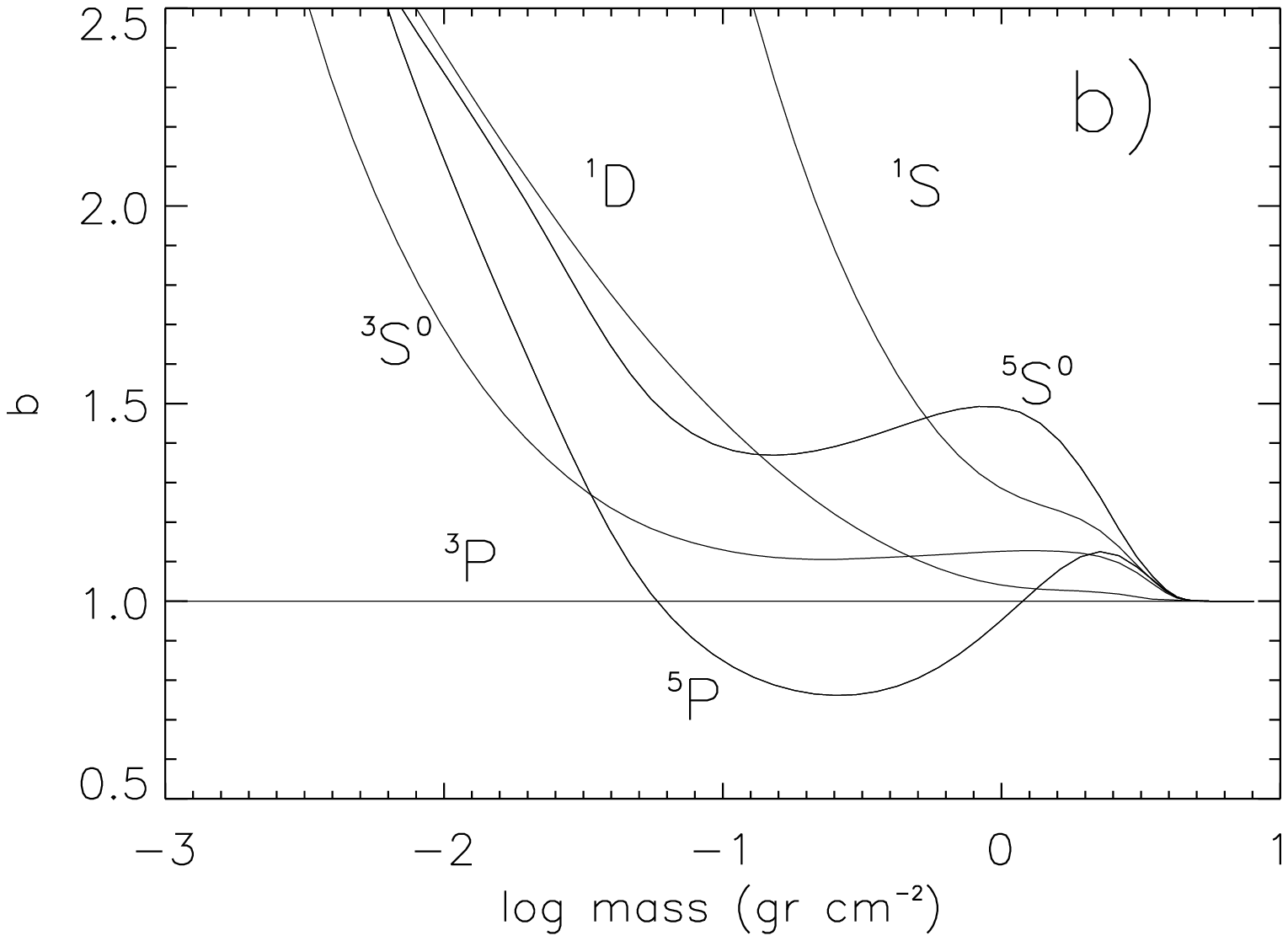}
\figcaption{
a) Grotrian diagram for O I (see Paper I). b) Departure coefficients
for the lowest 6 levels of OI.
\label{fig8}}
\end{center}
\end{figure*}

 Because of  the high excitation
 of the involved terms, one needs a very extended model atom to prevent
 spurious results in NLTE calculations of the triplet.
 Our model atom (54 levels and 242 radiative transitions) 
 and the departure coefficients for the lowest 6 levels 
 of O I are depicted in Fig. \ref{fig8}. The lower level of the oxygen 
 triplet ($^5$S$^{\rm o}$)
 is overpopulated, while the upper level ($^5$P) is underpopulated in
 the line formation region. Consequently,
the line source function drops below the Planck function. This,
together with an increased line opacity, will produce 
deeper profiles.  Our results
 are similar to the calculations described by Kiselman (1993) for a model atom
 with 44 levels. In contrast with the situation for the infrared triplet, 
 Fig. \ref{fig8} shows that the departures from LTE expected for the two
 lowest levels ($^3$P and $^1$D), 
which are connected by the [O I] line at 6300 \AA, are 
 small in the deep layers where this weak line is formed 
 ($\log \tau \simeq -1.0 \Leftrightarrow \log$ mass $\simeq 0.1$).

\begin{figure*}[bottom!]
\begin{center}
\includegraphics[width=10cm,angle=90]{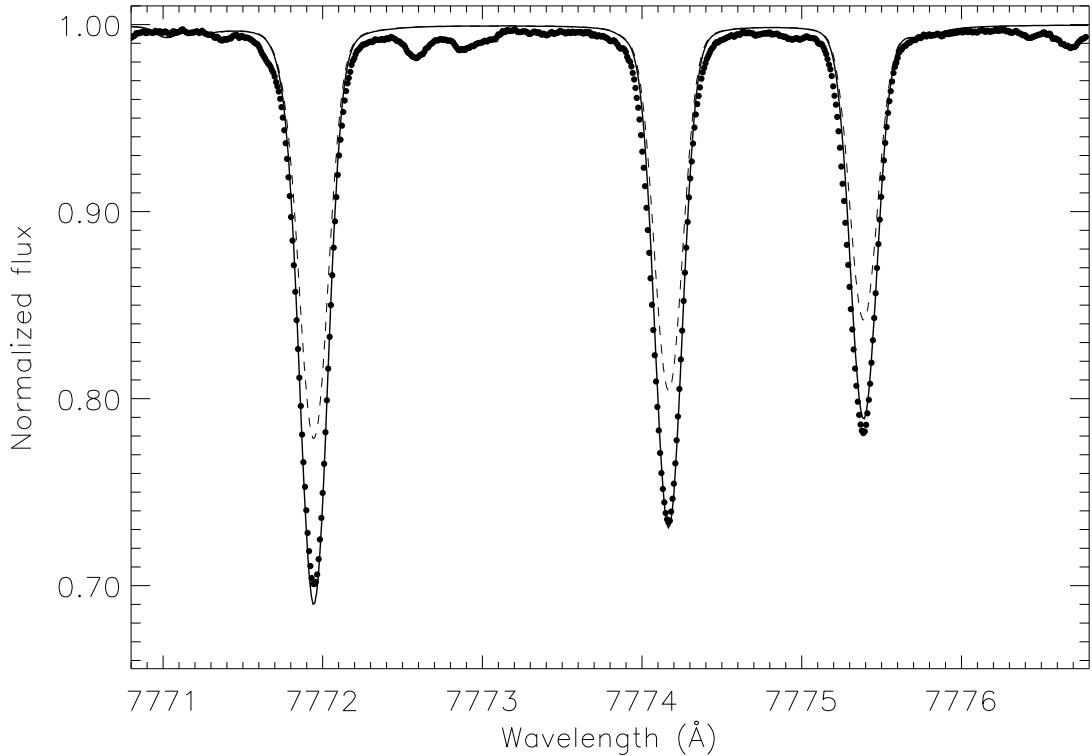}
\figcaption{
Comparison between the observed (dots) and calculated line profiles
for the O I infrared triplet. The solid line shows the NLTE calculation, and the
dashed line shows the LTE profiles.
\label{fig9}}
\end{center}
\end{figure*}

Our calculations 
 for the triplet give significant deviations from LTE, and indicate that 
 the observed lines are reproduced  with 
  $\log \epsilon\footnote{$\epsilon ({\rm O}) = 
  10^{12}$ N(O)/N(H), where N(E) is the number density of
  the element E.}({\rm O}) = 8.61$, 
  as shown in Fig. \ref{fig9}.  
  Allende Prieto, Lambert \& Asplund (2001b) 
 showed  that the  forbidden line 
 at 6300 \AA, which forms very close to LTE conditions, is blended with 
 a Ni I line, and considering both with a detailed 3D LTE hydrodynamical 
 model they derived $\log \epsilon({\rm O}) = 8.69 \pm 0.05$. 
 Kiselman \& Nordlund (1995) showed 
 that a two-level NLTE calculation in a 3D model atmosphere predicted 
 weaker lines than the Holweger-M\"uller model, and we believe 
 that differences between 1D and 3D are 
 the reason for the small remaining discrepancy. 
 Kiselman \& Nordlund (1995)
 find an average decrease of the total equivalent width of the triplet by
 about 17 m\AA\ when computed in 3D respect to the 1D value.  If we change the 
 abundance from 8.61 to 8.69 dex, the calculated 
 total equivalent width of the triplet increases by 18 m\AA. 
 
 A recent study of hotter (spectral types 
 A-B) stars by Przybilla et al. (2000) has suggested that collisions 
 with electrons as described by Van Regemorter's formula are significantly 
 overestimated. If such a problem applies also to solar-like stars, 
 our abundance could also decrease, drifting away from the value derived
 from the forbidden line. Introducing collisions with hydrogen 
 in the calculations might also 
  decrease the derived abundance. The LTE profiles require an abundance of
  $\sim \log \epsilon({\rm O}) = 9.0$  in order 
  to fit the observed profiles.

                            
 In short, our calculations seem to bring the abundance derived from the 
 oxygen IR triplet into agreement with that obtained 
 by Allende Prieto et al. (2001b) from the forbidden line at 6300 \AA. 
 
\subsection{Sodium}

\begin{figure*}[top!]
\begin{center}
\includegraphics[width=9cm,angle=0]{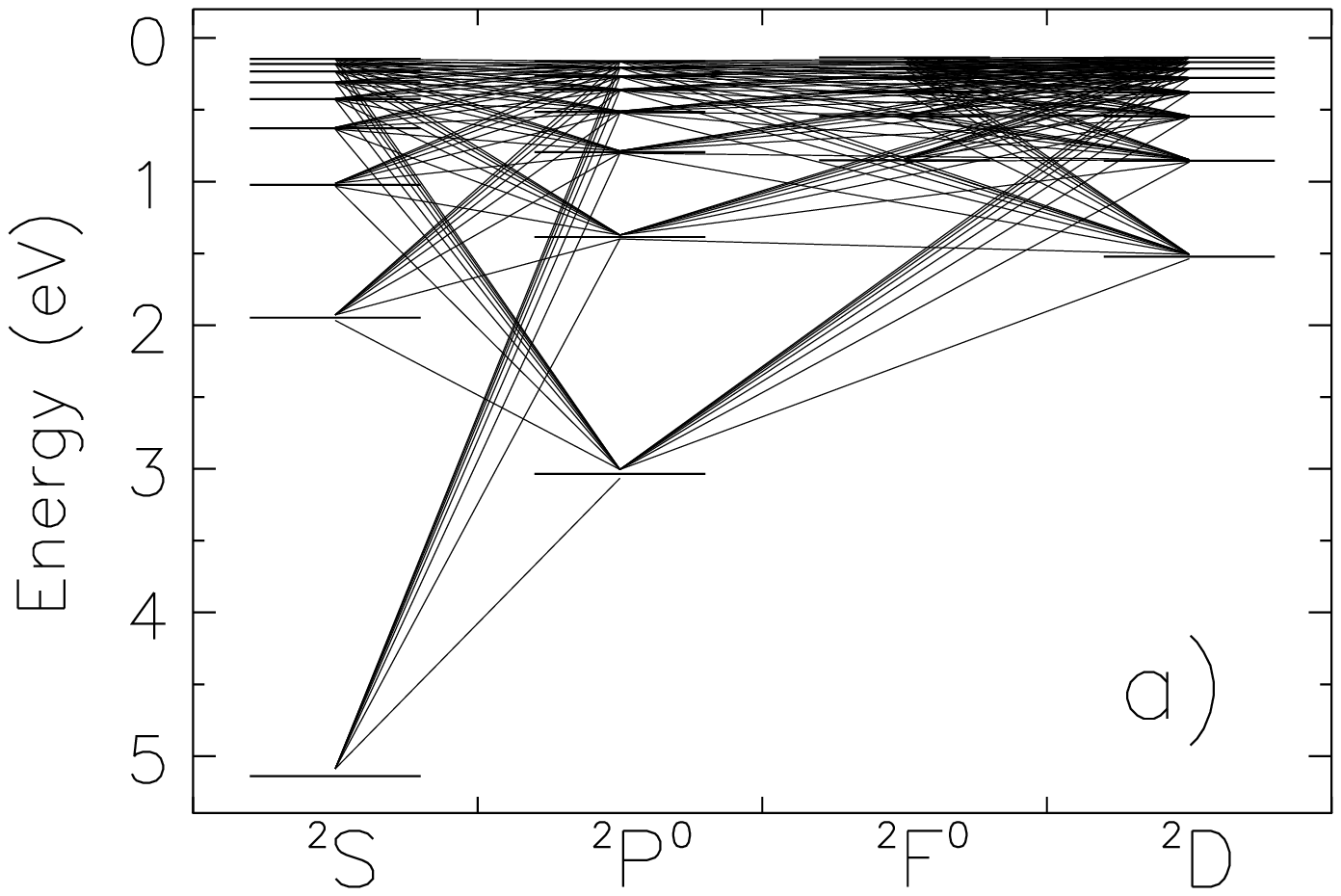}
\includegraphics[width=8cm,angle=0]{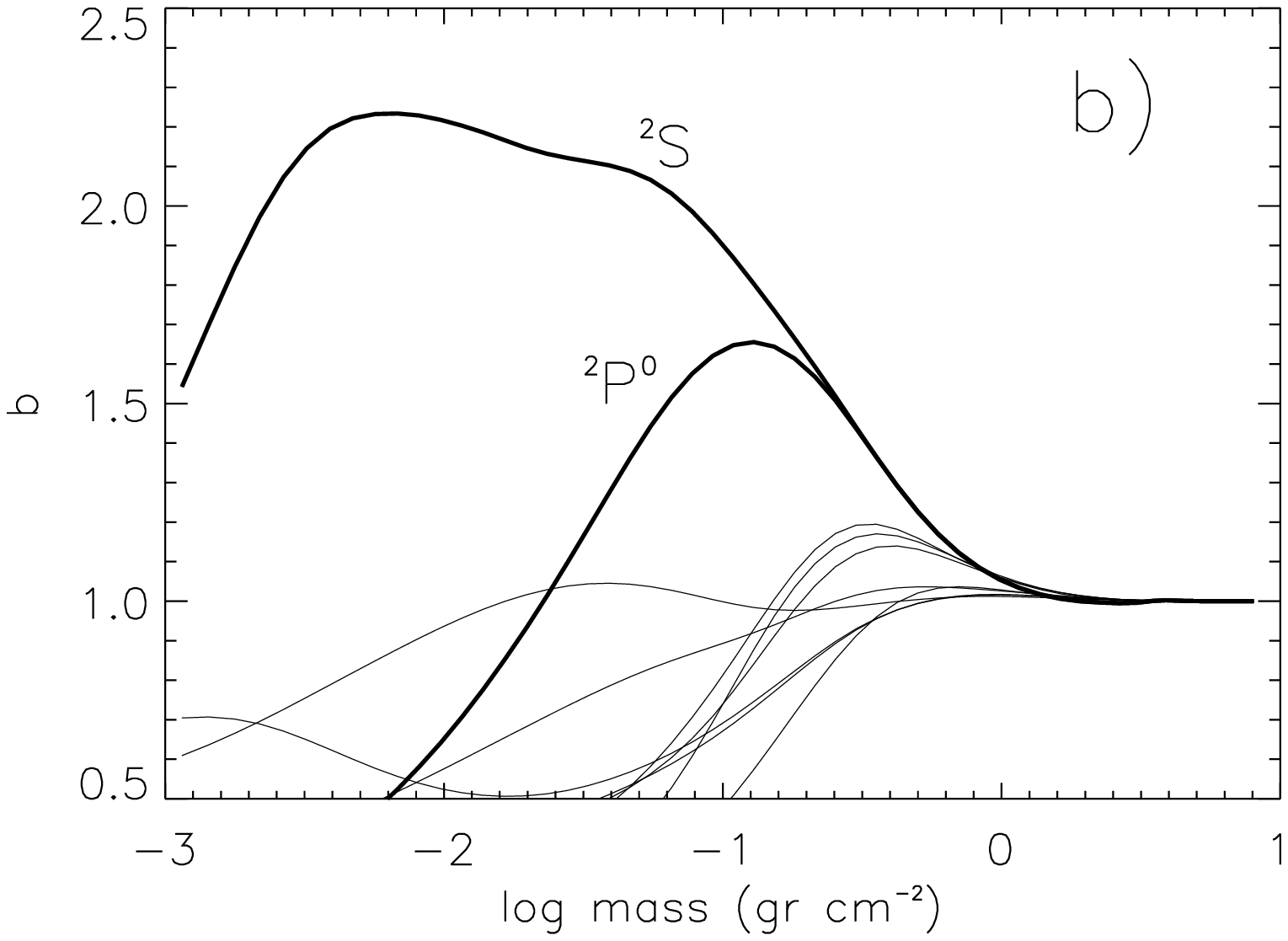}
\figcaption{
a) Grotrian diagram for Na I (see Paper I). b) Departure coefficients
for the lowest 10 levels.
\label{fig10}}
\end{center}
\end{figure*}

 We adopt Grevesse \& Sauval's (1998) photospheric 
 abundance,  $\log \epsilon ({\rm Na}) = 6.33 \pm 0.02$, which 
 is indistinguishable from the meteoritic abundance $6.32 \pm 0.02$ dex. 
Baum\"uller, Butler \& Gehren (1998) studied the NLTE 
formation of Na I lines in  
  solar and metal-poor atmospheres, and found a similar solar abundance,
   concluding that solar Na I profiles are best 
  reproduced including collisions with H atoms as prescribed by 
  Drawin (1968), but decreased by a factor of 20. 
 From these author's line list, we selected several lines that 
 appear to be clean in the solar   spectrum.

\begin{figure*}[top!]
\begin{center}
\includegraphics[width=12cm,angle=90]{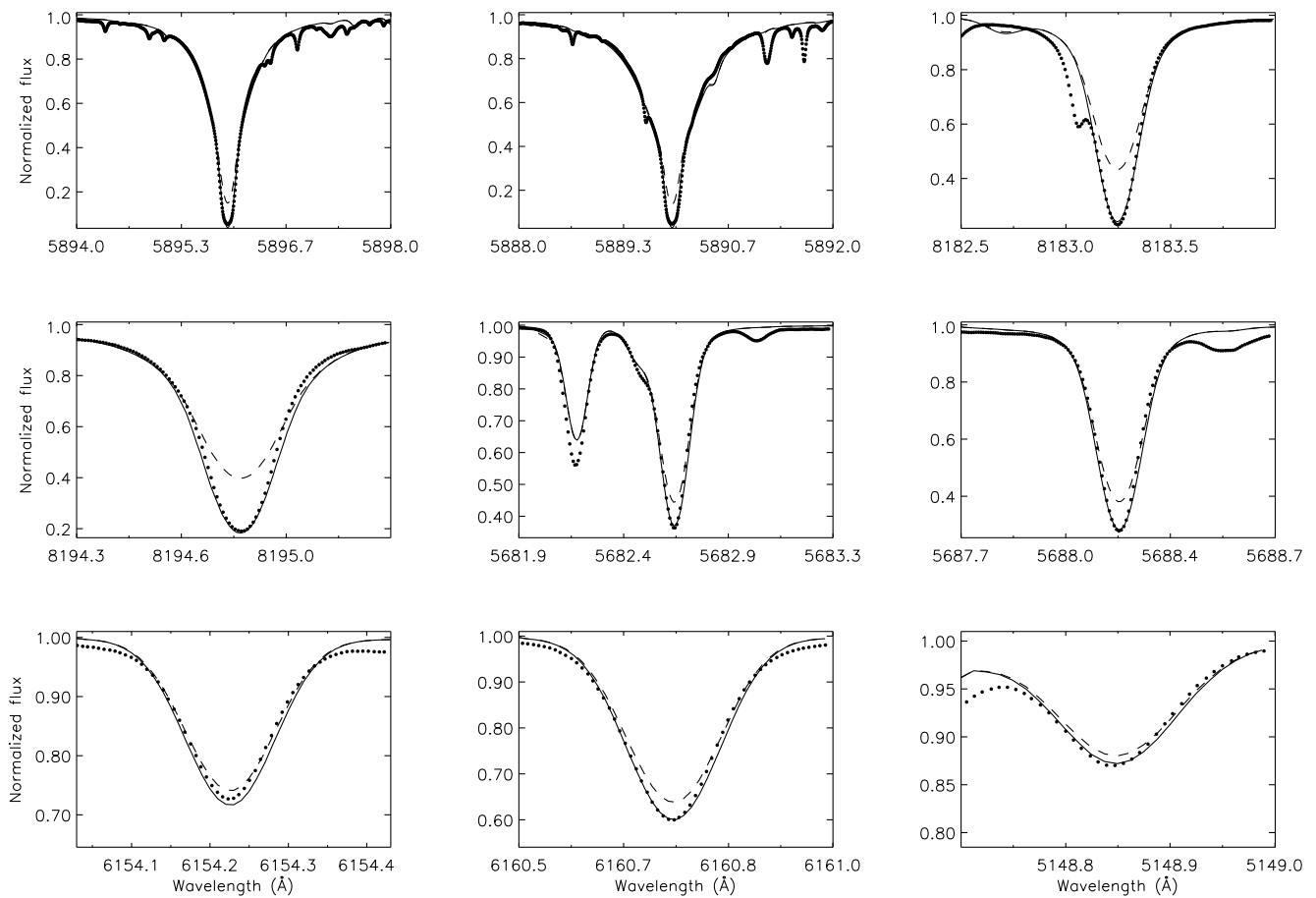}
\figcaption{
Comparison between the observed (dots) and calculated line profiles
for several Na I lines. The solid line corresponds to 
the NLTE calculation, and the
dashed line shows the LTE profiles.
\label{fig11}}
\end{center}
\end{figure*}

  Figure \ref{fig10} shows our Grotrian diagram for Na I (32 levels and 192 radiative
  transitions) and the departure coefficients
  for the lowest 10 levels. The two lowest levels, which are connected by the 
  D lines and have been identified in the Figure with thick lines, 
  are overpopulated in the photosphere compared to LTE. Most other
  levels are underpopulated. Our results are similar
  to those previously reported in the literature (see Bruls et al. 1992 for a
  extended discussion). It is evident from Figure \ref{fig11} that NLTE 
  improves significantly  the agreement with observations. 
  Our tests also show, as expected based on 
   \S \ref{uv}, that the NLTE calculations for Na I 
  do not depend much on the near-UV opacity.

 The Na D lines deserve some extra comments. The core of these lines 
 is formed in very high atmospheric layers, 
 and therefore our results for these lines should be taken with caution. 
 In spite of this problem, the overall agreement between our calculations
 and the observed D lines is satisfying.

\section{Conclusions and future work}

We have built a collection of model atoms  for use in 
NLTE calculations for late-type stellar atmospheres. The models 
 were compiled 
 in an attempt to use a series
 of data sources as homogeneous as possible. The
 details are presented in Paper I, and the data are formatted for the code 
  {\sc Tlusty} (Hubeny 1998; Hubeny \& Lanz 1995). 
  
In this paper, we test the data by computing the flux emerging from the
solar surface and comparing it to high-accuracy observations from space. 
Our calculations are based on a {\it fixed} atmospheric structure 
calculated in LTE, and the level populations are computed both in LTE and
in NLTE. The agreement with the observed fluxes is relatively good, especially
at wavelengths longer than about 2700 \AA, where the LTE computed fluxes tend to
be just slightly below the observations. 
In the region between 2400 and 2600 \AA, where the Mg I bound-free
opacity dominates over other sources, we find indication that the 
calculated LTE fluxes are larger than observed. The disagreement is even 
 slightly enhanced for the NLTE calculations. We note that a possible opacity 
 deficit in this window may have an impact on determinations of boron 
 abundances from the B I line at 2497 \AA.
 The NLTE  flux is systematically too high between 1700  and
 2100 \AA, where Al I photoionization plays a major role and the LTE 
 predictions match reasonably well the observations. 

 The good agreement we find between observed and predicted absolute fluxes
 apparently contradicts a previous result by Bell, Balachandran \& 
 Bautista (2001). With the data available to us, it is not possible to 
 clearly identify a single reason for the discrepant results. 
 Differences between the photoinization cross-sections for Mg I, Ca I, Si I, 
 and C I, but mainly Fe I, are expected. The computer codes
 used in the calculation are not the same -- e.g., H and H$^{-}$ bound-free 
 and free-opacities may differ slightly. The model atmospheres discussed
 may also be somewhat different. The only obvious factor that we can suggest
 is responsible for at least part of the differences is the choice of 
 abundances.  Bell et al. adopted $\log \epsilon({\rm Mg}) = 7.44$ and 
 $\log \epsilon ({\rm Fe)}= 7.55$. We adopted, 
 following Grevesse \& Sauval (1998), 
 $\log \epsilon({\rm Mg}) = 7.58$ and $\log \epsilon ({\rm Fe)} = 7.50$. 
 Grevesse \& Sauval quote an uncertainty of 0.05 dex for these photospheric 
 abundances. Their figures are also in perfect agreement with their meteoritic 
 values -- these only uncertain by 0.01 dex. 
 We approximately estimate the effect of lowering the Mg abundance 
 on the predicted flux as an increase of 14 \%, 7 \%, and 3 \% at
 2450, 2800 and 3000 \AA, respectively.

 LTE and NLTE calculated fluxes systematically depart from observations
 below $\sim 1700$ \AA, where Si I bound-free absorption becomes very 
 important. We expect the formation region for these fluxes 
 to be at very high atmospheric layers, where many of the assumptions
 involved in our calculations, in particular radiative equilibrium,
 may break down. 
 A model with a temperature minimum (e.g. Vernazza et al. 1976) will 
 surely perform much better in this spectral band.

 From the comparison between observed and calculated fluxes between 2600 and
 4000 \AA, we conclude  that there is no solid foundation for previous 
 claims of missing continuum opacity in the near-UV. 
 It has also become apparent that 
 departures from LTE will affect the shortest wavelengths, 
 below $\sim 2600$ \AA.  
Coupling  between the populations of different species that 
produce significant absorption in the UV is very likely, and detailed and
simultaneous calculations for all of  them are necessary. 

Our NLTE predictions
for the species that play a important role in the UV, such as Al I or Mg I, 
fail to match the observed optical and near-infrared 
line profiles. In contrast, 
the NLTE profiles for Na I or O I, which do not absorb significantly
in the UV, represent a clear improvement over their LTE counterparts. 
As our NLTE computations 
 involve adopting a {\it fixed} LTE atmospheric structure (electron pressure
 and temperature and gas pressure), it seems a likely possibility that the
adopted structure does not represent well the physical conditions in the
highest atmospheric layers and, in turn, the level populations predicted
in NLTE for UV-relevant species  are wrong. 
However, we should exercise caution, as previous studies have successfully
performed NLTE analysis of optical lines of 
neutral Al and Mg (Baum\"uller \& Gehren 1996; Zhao, Butler, \& Gehren 1998).
Replacing our statistical approach to account for the line absorption,
which is based on the concept of superlevels and opacity distribution
functions, has been shown to introduce only
mild differences for early-type stars (Lanz \& Hubeny 2001), but
may be crucial for solar-type stars. 
Consistent NLTE calculations to solve both
the atmospheric structure and the populations must be carried out to 
avoid possible systematic errors in the solution of the statistical
equilibrium for species such as Al I or Mg I.

An acceptable NLTE solution should reproduce satisfactorily 
the UV 
fluxes and line profiles in the UV, optical, and near-infrared domains, but
it is unclear that such goal is possible within the framework of plane-parallel,
homogeneous, 
static model atmospheres. Anderson (1989) computed a line-blanketed NLTE
model in hydrostatic and radiative 
equilibrium for the Sun, noting significant differences
in the atmospheric structure with respect to LTE for $\tau \lesssim 10^{-3}$.
Recently,  Hauschildt, Allard, \& Baron (1999) computed 
consistent NLTE  model atmospheres
for the Sun and Vega ($T_{\rm eff} \sim 10,000$ K), finding  small
differences between LTE and NLTE structures for the former, but important
for the latter. However, their calculations treated neither  Al I nor Mg I
in NLTE. 

Our statistical equilibrium calculations neglect collisions with 
hydrogen atoms.
Such omission does not prevent us from 
obtaining a
 good match of the observed line profiles of O I and Na I lines, but 
 it might have a more important effect on 
 Mg I and Al I (see, e.g., Baum\"uller \& Gehren 1996; 
 Zhao et al. 1998). Although our success in matching within a few percent  
 the solar absolute flux between 2700 and 4000 \AA\ is encouraging, validation
 of the atomic models presented in Paper I requires further investigation 
 considering a self-consistent NLTE calculation of the atmospheric structure, 
and  the effect of H collisions. Our LTE tests shown CO not to be very 
important to accurate model the solar UV flux, but this and other molecular 
species, in particular OH and CH (see Kurucz, van Dishoeck \& Tarafdar 1987), 
 should be properly considered in NLTE to secure this result. 
Finally, enlarging the set of available high-accuracy UV spectrophotometry 
by observing other nearby stars is necessary, as it is to  
explore the effect of granulation on the UV flux.

\acknowledgments

We thank N. Piskunov for assistance with VALD. NSO/Kitt Peak FTS data used here
were produced by NSF/NOAO. We have taken advantage of  VALD, ASD/NIST, and
NASA's ADS. This research has been partially
supported by the NSF through grant AST-0086321 and by the Robert A. 
Welch Foundation of Houston, Texas.

\end{document}